\documentclass[english,12pt, a4paper]{article}

\usepackage{amssymb,amsfonts,amsmath}
\usepackage{a4wide}
\usepackage{indentfirst}
\usepackage{graphicx}
\usepackage[round]{natbib}
\usepackage{styleEli}
\usepackage{lineno}

\usepackage{smallcaption}
\begin{document}

\begin{center}
\Large{\bf Forward and adjoint quasi-geostrophic models} 

\Large{\bf of the geomagnetic secular variation}

\myspace

\Large{Elisabeth Canet$^{1*}$, Alexandre Fournier$^2$, Dominique Jault$^1$}

\end{center}
\myspace

\normalsize{ $^1$\ Laboratoire de G\'eophysique Interne et Tectonophysique, Universit\'e Joseph-Fourier, CNRS, Grenoble, France}

\normalsize{ $^2$\ \'Equipe de G\'eomagn\'etisme, Institut de Physique du Globe de Paris Universit\'e Paris Diderot, INSU/CNRS, Paris, France}

\normalsize{ $^*$\ Corresponding author: Elisabeth.Canet@obs.ujf-grenoble.fr}

\myspace

\begin{center}
\normalsize{ Published in Journal of Geophysical Research, 2009, 114, B11, doi:10.1029/2008JB006189}

\myspace

\end{center}
\begin{abstract}
We introduce a quasi-geostrophic model of core dynamics, which aims at describing core processes
on geomagnetic secular variation timescales.
It extends the formalism of Alfv\'en torsional oscillations by incorporating non-zonal motions.
Within this framework,
the magnetohydrodynamics takes place in the equatorial plane; it
involves quadratic magnetic quantities, which are averaged along the direction of rotation of the Earth.
In addition, the equatorial flow is projected on the core-mantle boundary. It interacts with the
magnetic field at the core surface, through the radial component of the magnetic induction equation. 
That part of the model connects the dynamics and the observed secular variation, with
the radial component of the magnetic field acting as a passive tracer.
We resort to variational data assimilation to construct formally the relationship between model predictions
and observations.
Variational data assimilation seeks to minimize an objective function, 
by computing its sensitivity to its control variables.
The sensitivity is efficiently calculated after integration of the adjoint model.
We illustrate that framework with twin experiments, performed first in the case of the kinematic core flow inverse problem,
 and then in the case of Alfv\'en torsional oscillations.
In both cases, using the adjoint model allows us to retrieve core state variables which, while taking part in
the dynamics, are not directly sampled at the core surface.
We study the effect of several factors on the solution (width of the assimilation time window, amount and quality of data),  
and we discuss the potential of the model to deal with real geomagnetic observations.
\end{abstract}

\section{Introduction}
Current descriptions of core dynamics rely on two sources of information: observations of the magnetic field, and 
physical laws governing the evolution of the state of the core.
The Earth's magnetic field is assumed to have an internal origin through the process of geodynamo;
it is generated and sustained by fluid motions in the metallic liquid outer core, and  
varies on a wide range of time scales reflecting the various time and space
scales of core magnetohydrodynamics.

The quality of observations of the Earth's magnetic field has much improved 
since the set-up of the first network of magnetic observatories by Gauss and 
co-workers in $1834$, which was followed by the large increase in the number of observatories 
at the beginning of the twentieth century. Other turning points have occurred since: 
the introduction of the proton precession magnetometer, the development of declination/inclination 
magnetometers (DIflux) widely used in observatories by the $1970$'s, and finally the 
rise of the Intermagnet network of digital observatories sharing modern measurement practices after $1990$ 
\citep[see, e.g., the review by][]{TOG05_04}. 
The good temporal coverage of observatory data has now been supplemented by the excellent spatial coverage of satellite data.
Following the launch of three low Earth orbiting satellites -Oersted, CHAMP and SAC-C-
supplying geomagnetic data, a continuous satellite time series extends now to $10$ years. 

The magnetic field can be downward continued throughout the solid mantle to the fluid core surface.
Field models are built, describing the radial component of the 
main field and its time evolution at the core-mantle boundary (CMB) \citep{TOG05_02,TOG05_05}.
Calculating the geomagnetic secular variation, the first time derivative of the main field time series,
emphasizes rapid changes of the magnetic field, on characteristic time scales ranging from years to centuries.
Inversions of a snapshot of the geomagnetic secular variation can be performed using the radial induction equation at the CMB, 
in order to retrieve the large-scale part of the flow beneath it 
\citep{eymin2005,holme2006,pais2008,olsen2008}. 
The root mean square (rms) speed of these flows is typically of the order on $15$~km/y. 
Such inversions, however, face non-uniqueness problems \citep{backus1968}.
Further assumptions are thus required to remove the non-uniqueness, 
and a great part of the work consists in finding constraints and regularizations to specify the  
flow \citep[section 8.04.2]{TOG08_04}. 
Alongside these kinematic inversions, numerical models of the geodynamo have been available for more than $10$ years, 
since the pioneering work of \cite{glatzmaier1995}. 
The magnetic field generated by those dynamical models explains features of the Earth's magnetic field (dipolar geometry, spatial spectrum); 
yet their parameters
are far from those of the Earth's core \citep[section 8.08.4]{TOG08_08}.
\cite{rau2000} and \cite{amitOlsonChristensen2007} tried to connect these two approaches (core flow inversion and forward numerical modelling) 
when inverting synthetic data from dynamo models. They found their core flow inversion method and the additional regularization
to be adequate for the retrieval of large-scale flow and magnetic field patterns.

A quality-control of core flow models is the angular momentum they carry \citep{jault88}.
Comparison of these estimates with core angular momentum changes inferred from decadal length-of-day variations is encouraging,
yet discrepancies remain. 
Angular momentum series are also derived from  atmospheric and oceanic flow models, based upon 
the data assimilation methodology \citep{kalnay1996}.
Those variations of angular momentum account very well for the observed seasonal and 
interannual changes in length-of-day \citep[section 3.09.4]{chen2005, TOG03_09}.

Data assimilation, routinely used in atmospheric science and more recently in oceanography, 
is now in early stages of use in the field of core physics.
Applied to the core, this technique should allow us to interpret the secular variation in terms of dynamics, 
thereby enlarging the work done on kinematic core flow inversion.
Resorting to a toy model, \cite{fournier07} assimilated synthetic data in a one-dimensional model that retains characteristic features
of the induction and Navier-Stokes equations. They concluded that a good knowledge of the 
observed magnetic field can be translated 
into a good knowledge of core flow, through the
process of data assimilation, which takes explicitly into account the dynamical relationship that exists 
between magnetic and velocity fields. 
This conclusion was also drawn by \cite{sun2007}, using a much similar toy model and a different implementation 
of data assimilation (sequential as opposed to variational). 
In a preliminary study, \cite{liu2007}  applied an optimal 
interpolation scheme to a three-dimensional dynamo model, using
a synthetic set of observations. They showed in particular that
assimilation of observations could partially alleviate the negative
impact of wrong model parameter values. Still, the values of
the parameters used typically in that class of simulations are far
from being Earth-like, due to the numerical cost of their integration.
There is hope, however, that systematic and appropriate
exploration of the parameter space of those models will eventually
yield scaling laws of the kind proposed by \cite{christensen2006}, which will ultimately permit a 
reliable extrapolation between their output and the observed
secular variation. In the context of geomagnetic data assimilation,
the current situation is even worse, though, since one assimilation run requires
several tens of forward realisations.

A solution to this conundrum is to construct a simplified dynamical model, tailored to the study of the secular variation.
In this paper, we introduce a model which relies on quasi-geostrophic dynamics. 
As a matter of fact, on rapid time scales as those characterizing the secular variation, 
rotational forces prevail on magnetic forces in the bulk of the fluid \citep{jault2008}.
The resulting two-dimensional flow interacts with the radial component 
of the magnetic field (in that instance, a passive tracer) at the core-mantle boundary.
The quasi-geostrophic assumption has recently been used in the framework of kinematic core flow inversions  \citep{pais2008,gillet2009}. 
It also provides us with the tools necessary to build a dynamical model of the geomagnetic  secular variation.

Being quasi-geostrophic, this model comprises the equations for torsional Alfv\'en waves, 
for which the dynamics is axisymmetric; Alfv\'en torsional
waves are associated with geostrophic (zonal) motions in the core
\citep{braginsky1970}. The frequency of these oscillations is proportional 
to the rms strength of the magnetic field $\mags$ perpendicular to the rotation axis ($s$ denotes the cylindrical radius).
Accordingly, using a database of computed core flows, \cite{zatman97,zatman99} and \cite{buffett2009} 
have calculated radial profiles of the quadratic cylindrical radial component of the magnetic field averaged on geostrophic cylinders, 
$\axiBsBs$, within the core.

Our quasi-geostrophic model generalizes that axisymmetric approach by adding non-zonal motion and magnetic field.
Theoretical solutions of the model include various families of diffusionless hydromagnetic waves, some of which were first studied by 
\cite{hide1966} in order to explain the observed secular variation.

The goal of this paper is thus to describe a quasi-geostrophic 
forward model of the Earth's core fast dynamics, and to place it at the heart of a
geomagnetic data assimilation process. 
In section 2, we derive that quasi-geostrophic model, along with its link to the observations at the CMB. Variational data assimilation
is introduced in section 3 and its principles are illustrated in section 4 with twin experiments. 
That section begins with the study of 
the classical kinematic inversion of a steady core flow, set within that framework. 
A second illustration is dedicated to the retrieval of the magnetic field sheared by
Alfv\'en torsional waves. Results are summarized and discussed in section 5.

\section{Quasi-geostrophic forward model}
\label{2Dmodel}

\begin{figure}
\centerline{\includegraphics[width=0.95\linewidth]{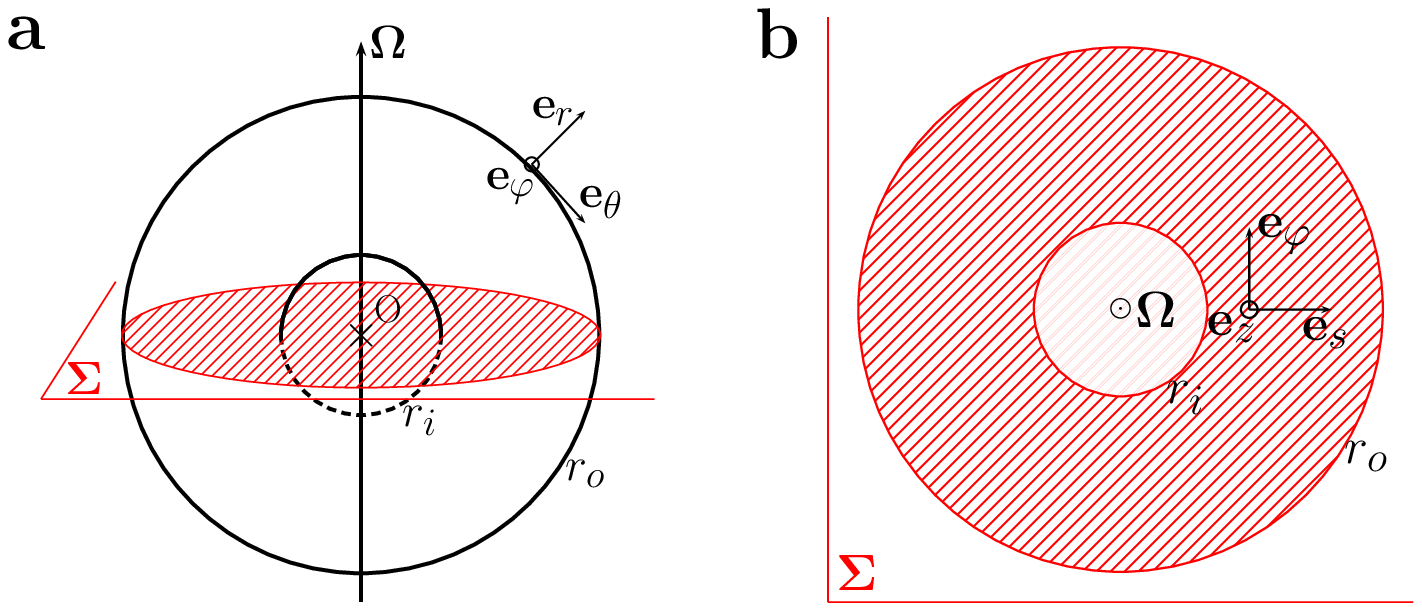}}
\caption{Geometry of the system and notations; a: Side view, b: Equatorial section. 
$\Sigma$ is the equatorial plane and the CMB is the outer sphere, located at $r=r_o$. $r_i$ is the radius of the inner core. Spherical 
(O,~$\vectr, \vecttheta, \vectp$) and cylindrical (O,~$\vects, \vectp, \vectz$) 
coordinate system bases are defined at the core surface and in $\Sigma$ respectively.}
\label{geom}
\end{figure}

We shall model the Earth's outer core as a spherical fluid shell 
of inner radius $r_i$ and outer radius $r_o$.
The fluid has density $\rho$, and it is electrically conducting. Its magnetic 
diffusivity is $\eta$. 
The system is rotating at angular velocity $\Omega$ around the $z$-axis.
Figure~\ref{geom} sketches the  geometry and summarizes the notations.

The fluid is characterized by its magnetic field $\magn$, its velocity $\velo$, and the reduced pressure $\Pi$, 
that includes pressure and the centrifugal potential.
We choose $r_o$ as length scale and $B_0$, a typical magnetic field intensity in the core interior, as the magnetic field scale.
Velocities are scaled with the Alfv\'en waves speed 
\begin{linenomath*} 
\begin{eqnarray}
V_A = \frac{B_0}{\sqrt{\mu_0 \rho}},
\end{eqnarray}
\end{linenomath*} 
in which $\mu_0$ is the magnetic permeability of free space.
The pressure scale is $\rho V_A^2$, and the time scale is the Alfv\'en waves period: $T_A=r_o/V_A$.

The evolution of the magnetic field in the core is governed by the induction equation, which has the dimensionless form,
\begin{linenomath*} 
\begin{eqnarray}
\ddt \magn  = \mathbf \rot \left( \velo \times \magn \right) + {S}^{-1} \mathbf \nabla  ^2 \magn. \label{indeq}
\end{eqnarray}
\end{linenomath*} 

The Lundquist number $S$ characterizes the ratio between the magnetic diffusion time and the period of Alfv\'en waves
\citep[e.g.][]{roberts1967},
\begin{linenomath*} 
\begin{eqnarray}
S=\frac{r_o V_A}{\eta}.
\end{eqnarray}
\end{linenomath*} 
For the Earth's core, $S$ is of the order of $10^{4}$ to $5 \times 10^{4}$ \citep{jault2008}.
On secular variation time scales, diffusion 
becomes negligible compared to induction, hence the large value of $S$, 
which yields the frozen-flux approximation. 

Assuming that the mantle is electrically insulating on secular variation time scales, the magnetic field can be downward continued 
to the core-mantle boundary. 
In the case of a perfectly conducting fluid (the frozen-flux limit), the radial component of the magnetic field $\magr$ is the sole magnetic component
continuous across the spherical CMB.
At the top of the core, $\magr$  interacts with core motions by means of
the radial component of the diffusionless version of equation~(\ref{indeq}) at $r=r_o$, 
\begin{linenomath*} 
\begin{eqnarray}
\ddt \magr = - \divH \left( \velo \magr \right),  \label{tracerbr}
\end{eqnarray}
\end{linenomath*} 
with the horizontal divergence operator $\divH $ defined as
\begin{linenomath*} 
\begin{eqnarray}
\divH {\bf v} = \left( \sin \theta \right)^{-1} \ddtheta \left( \sin \theta  v_{\theta} \right) + \left( \sin \theta \right)^{-1} \ddp v_{\varphi}, 
\end{eqnarray}
\end{linenomath*} 
where $(r, \theta, \varphi)$ are the spherical coordinates.
It is that equation at the core-mantle boundary that connects our model to the observations. 
The time-varying $\magr$ acts as a passive tracer (a drifting buoy), because
it interacts with the velocity field at the core surface  and does not affect the 
dynamics that sets up in the interior of the core (see below).

On secular variation time scales rotation forces are much larger than magnetic forces in the bulk of the fluid. 
\cite{jault2008} suggests that rapidly rotating motions of lengthscale $l$ are axially invariant if the non-dimensional 
Lehnert number, $\lambda_{l}$, is small enough.  
That number measures
 the ratio between the period of inertial waves, $1/\Omega$, and the period of Alfv\'en waves,  
$l/V_A$ \citep{lehnert1954}:
\begin{linenomath*} 
\begin{eqnarray}
 \lambda_l = \frac{B_0}{\Omega(\mu_0\rho)^{1/2}l}\ \ \  ;\ \ \  \lambda_{r_o} = \frac{1}{\Omega T_A}, \label{lehnert}
\end{eqnarray}
\end{linenomath*} 
Note that  $\lambda_{l}$ is a decreasing function of $l$.
In his calculations, the flow appears to be invariant in the direction parallel to the rotation axis, provided $\lambda_l \ll 1$. 
For the Earth's core, with $B_0$ of the order of $2$~mT  \citep{christensen2009}  and $l \approx 10^6$ m, $\lambda_l \approx 10^{-4}$.
%
Therefore, we shall assume that the flow is geostrophic at leading order.
Working in the equatorial plane $\Sigma$ (crosshatched in Figure~\ref{geom}), a cylindrical set of coordinates 
$(s, \varphi, z)$, with $\vectz$ parallel to the axis of rotation,
is well-suited to study the resulting columnar patterns.

The main force balance involves  the Coriolis force and the pressure gradient
\begin{linenomath*} 
\begin{eqnarray}
2 \vectz \times \velo^0 = -\mathbf \nabla   \Pi^0, \label{geos}
\end{eqnarray}
\end{linenomath*} 
where the superscript $0$ denotes the main order.
Taking the curl of equation~(\ref{geos}) yields the Proudman-Taylor theorem, namely the $z$-invariance of the 
flow.

Within a spherical container, $\velo^0 $ does not satisfy the non-penetration boundary condition
at the CMB, except if it consists of cylindrical flows
organized around the rotation axis.
Thus we have to add the  first-order contribution in $\lambda_{r_o}$ of the Coriolis force, leading to
\begin{linenomath*} 
\begin{eqnarray}
\mathrm{D}_t \velo^0 + 2 \lambda_{r_o} ^{-1} \vectz \times \velo^1
= - \boldnabla   \Pi^1 +  \left( \rot \magn  \right) \times \magn, \label{NS_1}
\end{eqnarray}
\end{linenomath*} 
where $\mathrm{D}_t$ denotes the material derivative $\mathrm{D}_t = \ddt + \Xgrad{\velo^0}$. 
At first-order, magnetic forces are a natural candidate to trigger a departure from geostrophy, since
magnetic energy is large compared to kinetic energy in Earth's core. Buoyancy forces
are another candidate that we could additionally take into account, which
we discard for now for the sake of simplicity.
Viscous forces are neglected, while equation~($\ref{NS_1}$) shows that the Coriolis force is scaled with the inverse
of the Lehnert number $\lambda_{r_o}$.
The non-penetration boundary condition at the CMB: $\velo^1 \cdot \vectr = 0  \mbox{ at } z=\pm h$,yields a linear dependence of $\velo^1$ with respect to $z$
\begin{linenomath*} 
\begin{eqnarray}
\velz^1 (s>r_i, \varphi, z) =  z \beta \vels^0 (s, \varphi). \label{equz1}
\end{eqnarray}
\end{linenomath*} 
If $h = \sqrt{r_o^2 - s^2}$ denotes the half-height of the column located at a given cylindrical radius $s$,
the slope of the upper surface is ${\mathrm d}h/{\mathrm d}s$, and we can define
\begin{linenomath*} 
\begin{eqnarray}
\beta(s) = h^{-1} {\mathrm d}h/{\mathrm d}s.
\end{eqnarray}
\end{linenomath*} 
The notation  $\beta$ has been chosen in reference to the $\beta$-plane approximation. 
This approximation -with uniform $\beta$- is ubiquitous in geophysical fluid dynamics \citep[e.g.][section~2.3]{vallis06}.
It is convenient, indeed, to  study planetary Rossby waves assuming that
the Coriolis parameter ($f_0 = 2 \Omega \cos \theta$)
varies linearly with latitude; $\beta$ is then the northward gradient of the Coriolis parameter.

According to our quasi-geostrophic approach, the flow in the outer core is nearly two-dimensional, which makes it natural to
take the vertical average of the Navier-Stokes equation~(\ref{NS_1}).
The vertical average $\left\langle \cdot \right\rangle$ of a quantity $X$ is defined as
\begin{linenomath*} 
\begin{eqnarray}
\left\langle X \right\rangle \left( s, \varphi \right) = \intz{X}. \label{moyvert}
\end{eqnarray}
\end{linenomath*} 

In a multiply-connected domain, the $\varphi$-averaged vorticity equation is not equivalent 
to the  $\varphi$-averaged Navier-Stokes equation \citep{plaut2003}, as the former does not ensure 
the existence of the pressure field; 
accordingly, we describe the evolution of the non-zonal flow  ${\NZ{\velo}}$ by means of  the axial vorticity equation,
while the $\varphi$-averaged momentum
equation directly provides us with the time changes of the zonal velocity  $\velo^Z=\velp^Z \vectp$. 
In the remainder of this paper, the superscript capital $Z$ marks zonal quantities. 
It should not be confused with small $z$, which  refers to the direction of rotation.

The non-zonal ($NZ$) velocity field ${\NZ{\velo}}$ is written as the curl of a $z$-independent non-zonal
streamfunction $\Psi$,  
\begin{linenomath*} 
\begin{eqnarray}
{\NZ{\velo}} (s, \varphi) = \rot \Psi (s, \varphi)\vectz.  
\label{defpsi}
\end{eqnarray}
\end{linenomath*} 
The non-zonal vorticity field $\vort$ is defined by $\mathbf \vort = \rot {\NZ{\velo}}$, and its vertical component is
\begin{linenomath*} 
\begin{eqnarray}
\vortz (s, \varphi)= - \laplE \Psi(s, \varphi),
\end{eqnarray}
\end{linenomath*} 
in which the equatorial Laplacian operator is defined by
\begin{linenomath*} 
\begin{eqnarray}
\laplE = s^{-1} \dds \left( s \dds \right) + s^{-2} \ddpp.
\end{eqnarray}
\end{linenomath*}

If we now curl the non-zonal part of the $z$-averaged Navier-Stokes equation~(\ref{NS_1}), we find that the vertical component 
of the vorticity equation is then identical to equation~($17$) of \cite{pais2008}, with an explicit right-hand side term,
\begin{linenomath*} 
\begin{eqnarray}
\mathrm{D} _t \vortz -  2  \lambda_{r_o} ^{-1} \beta s^{-1} \ddp \Psi &=& \left( s^{-1}\dds \ddp +s^{-2}\ddp \right) \left( \BpBp - \BsBs \right)
\nonumber \\ &&
+ \left(3s^{-1} \dds - s^{-2}\ddpp + \ddss \right) \BsBp. \label{eqvort}
\end{eqnarray}
\end{linenomath*} 

The magnetic surface terms, which appear 
 when taking the $z$-average of the Lorentz force, are neglected 
because we assume the magnetic field at
the core surface to be much smaller than in the bulk of the fluid.
The non-penetration boundary condition, at $s=r_o$ and at the tangent cylinder $s=r_i$, imposes $\Psi = 0$ at both boundaries.

The time evolution of the zonal velocity $\velp^{Z}(s)= s \omeg (s)$ is governed by 
\begin{linenomath*} 
\begin{eqnarray}
\mathrm{D} _t \omeg =  \left(s^3h \right)^{-1} \dds \left( s^2h \BsBp  \right). \label{eqzon}
\end{eqnarray}
\end{linenomath*} 

The two flow equations~(\ref{eqvort}) and (\ref{eqzon}) 
contain 
$z$-averaged squared magnetic quantities $\BsBs$, $\BpBp$ and $\BsBp$, whose 
time evolution is in turn derived from the diffusionless version of the induction equation~(\ref{indeq})
\begin{linenomath*} 
\begin{eqnarray}
\ddt \BsBs &=& -\left[ \velo^0 \cdot \nabla \right]\BsBs + 2 \BsBs \dds \vels^0 + 2s^{-1}  \BsBp \ddp \vels^0, \label{eqbs2}\\ 
\ddt \BpBp &=& -\left[ \velo^0 \cdot \nabla \right]\BpBp - 2 \BpBp \dds \vels^0 + 2 s \BsBp \dds \left(s^{-1} \velp^0 \right), \\
\ddt \BsBp &=&  -\left[ \velo^0 \cdot \nabla \right] \BsBp + s \BsBs \dds \left(s^{-1} \velp^0 \right)  + s^{-1} \BpBp \ddp \vels^0, \label{eqbsbp}
\end{eqnarray}
\end{linenomath*} 
where we have made use of the solenoidal character of $\magn$ and $\velo$.

The vertical averaging naturally sets the magnetohydrodynamics in the equatorial plane $\Sigma$ (Figure~\ref{geom}b).
The flow is then projected at the CMB, 
where it interacts with the radial magnetic field $\magr$ via equation~(\ref{tracerbr}) above. 

An alternative model, where the velocity field entering the set of equations~(\ref{eqbs2}) to (\ref{eqbsbp}) has a $z$-component given by (\ref{equz1})
 and a $\varphi$-component modified in order to ensure that the total velocity field remains solenoidal, is discussed in appendix~\ref{variante}.

\section{Variational data assimilation framework}
\label{framework}
In this section, we introduce the geomagnetic secular variation 
data assimilation problem with the notations suggested by \cite{ide97}. 
In comparison with the 4DVar label commonly used in data assimilation
\citep[e.g.][]{courtier1997}, our framework is 1+2DVar, since state variables
are defined in two-dimensional spaces to which a third (temporal)
dimension is added - the 3DVar label is traditionally reserved for
three-dimensional (in space) static problems \citep{courtier1997}.

The state vector $\etat$ for the Earth's core gathers the variables involved
in the description of the system state
\begin{linenomath*} 
\begin{eqnarray}
\etat  = \left[ \omeg, \Psi, \BsBs, \BpBp, \BsBp, \magr \right]^T,
\end{eqnarray}
\end{linenomath*}
where superscript $T$ means transpose.
Observations $\obs$ are available at $T_{\obs}$ different epochs and $N_{\obs}$ spatial locations
during the assimilation time window $\left[ 0, T\right]$; the size of $\obs$ is typically smaller than the dimension of $\etat$.
The observation vector is related to the true core state $\etat ^t$ via the observation operator $\obsope$:
\begin{linenomath*} 
\begin{eqnarray}
\obs = \obsope \etat ^t + \obserr,
\label{phys2obs}
\end{eqnarray}
\end{linenomath*}
in which $\obserr$ is the observation error. 

Variational data assimilation aims at adjusting a model solution $\forecast{\etat}$ to the observations \citep{talagrand97},
by minimizing a misfit function which comprises the quadratic discrepancy -if $\obsope$ is linear and errors are Gaussian- 
between the true observations and those predicted by  the computed state, $J_H$ \citep{courtier1997}:
\begin{linenomath*} 
\begin{eqnarray}
J_H = \sum_{j=1}^{T_{\obs}} \left[ \obsope_j \forecast{\etat}_j - \obs_j \right]^T \obserrmat_j^{-1}  
\left[ \obsope_j \forecast{\etat}_j - \obs_j \right],
\end{eqnarray}
\end{linenomath*}
where $j$ is the discrete time index and $\obserrmat = \statmean \left( \obserr \obserr^T \right) $ 
is the observation error covariance matrix, $\statmean (\cdot)$ denoting statistical expectation.
The matrix $\obserrmat$ describes the level of confidence we set in the observations.

It might be necessary to constrain the solution sought in order to enforce its uniqueness, especially if the problem is non-linear.
Constraining the assimilation refers to either adding a background state $\etat^b$ from which the estimate shall not strongly deviate, or 
applying additional constraints on the core state. 
Imposing a penalty, the goal of which is to favor a solution with a moderate level of complexity \citep[e.g.][]{courtier1987}, 
described by a matrix ${\sf C}$ applied to the state vector, 
consists in adding a second term to the objective function, of the form  
\begin{linenomath*} 
\begin{eqnarray}
J_C = \sum_{j=0}^{T} \etat_j ^T \const_j \etat_j, \label{regularizationterm}
\end{eqnarray}
\end{linenomath*}
where $\const={\sf C}^T{\sf C}$.
The total misfit function $J$ to minimize is then given by the sum
\begin{linenomath*} 
\begin{eqnarray}
J = \frac{\alpha_H}{2}J_H + \frac{\alpha_C}{2}J_C,\label{obsmisfit} 
\end{eqnarray}
\end{linenomath*}
in which the two contributions are weighted 
by two scalars, $\alpha_H$ and $\alpha_C$ \citep{fournier07}.

In practice, $\etat^f$ is the solution of a numerical non-linear model $\modelnl$, that describes the temporal 
evolution of $\etat$ at any discrete time $t_j \in \left[0,T\right]$,
\begin{linenomath*} 
\begin{eqnarray}
\etat_{j+1}^f = \modelnl_j \etat_{j}^f; 
\end{eqnarray}
\end{linenomath*}
that notation symbolically summarizes the set of equations developed in section~\ref{2Dmodel}.
Since the temporal history of the state is constructed with a dynamical model that couples its various components, 
we can, in principle, retrieve information about every state variable, 
even if only part of the state vector is directly probed by the observations on hand \citep{fournier07}.
Moreover, the initial condition $\etat _0$, termed the control vector, uniquely defines the model trajectory in state space.
That reduces the dimension of the associated inverse problem: we only seek the initial condition, $\etat_0$, 
starting from which the temporal evolution $\etat ^a$ 
will best fit the observations; in assimilation parlance, this best solution is called the analysis.
The minimization of $J$ (that is the search for  $\etat ^a$) is performed with a descent algorithm that involves the sensitivity
of $J$ to its control variables, $\etat_0$: $\nabla_{\etat_0} J$.
Its transpose 
is efficiently estimated with the use of the adjoint model $\modelnl^T$ \citep{ledimet86}.
For a given  $\etat _0$, one couples a forward integration of  $\modelnl$ with a
backward integration of $\modelnl^T$ to express the gradient of $J$. 
The adjoint model is that of the local tangent linear equations \citep[see e.g.][]{talagrand87,giering98}. 
Introducing the adjoint variable $\ADetat$ of $\etat$, the adjoint equation imposed by the cost function (equation~(\ref{obsmisfit})) is \citep{fournier07}
\begin{linenomath*} 
\begin{eqnarray}
\ADetat_{j-1} = \modelnl^T_{j-1} \ADetat_{j} + \alpha_H \obsope^T_{j-1} \obserrmat _{j-1}^{-1} 
\left(\obsope_{j-1} \etat_{j-1} - \obs_{j-1} \right) 
+ \alpha_C \const_{j-1} \etat_{j-1}, \label{eqadjoint}
\end{eqnarray}
\end{linenomath*}
where $\obsope ^T$ is the transpose of the observation operator (equation~(\ref{phys2obs})), which projects a vector 
from observation space to state space.
Through equation~(\ref{eqadjoint}), the adjoint field is fed by observation residuals (the difference $\obsope \etat - \obs$), 
as soon as there is an observation available.

The backward integration requires the knowledge of the model
trajectory over the assimilation time window. The storage of the
complete trajectory may cause memory issues, which are traditionally
resolved using a checkpointing strategy. The state of the system is
stored at a limited number of discrete times, termed checkpoints. Over
the course of the backward integration of the adjoint model, these
checkpoints are used to recompute local portions of the trajectory
on-the-fly, whenever those portions are needed \citep[e.g.][]{hersbach1998}.

Non-linear optimization is performed with the M1QN3 routine \citep{gilbert89}, 
which implements a limited memory quasi-Newton algorithm that approximates the inverse Hessian (second derivative) of $J$.

\section{Applications}

We show two illustrations of variational data assimilation applied to fast core dynamics, 
as described by our quasi-geostrophic model, with synthetic data.
The methodology of twin experiments is explained and applied to a steady non-zonal flow problem and, in a second
step, to a dynamical zonal model of torsional Alfv\'en waves.
These two problems correspond to two subsets of the model presented in section \ref{2Dmodel}.

\subsection{Twin experiments}
Instead of being satellite or observatory data, observations in our twin experiments are created from 
a synthetic true state, which is the result of the integration of the forward model for a given set of initial conditions, $\etat_0^t$.
Synthetic data have the advantage of representing only the physics involved in the model and are, in a first step, 
appropriate to test the implementation of the variational data assimilation algorithm.
A database of observations is produced with equation~(\ref{phys2obs}).

To construct the observation catalog, we include some geophysical realism by averaging the state at the core-mantle boundary. We choose the averaging window so that it corresponds
 to the ignorance of the spherical harmonic coefficients of degree $n > L$.
Then, the product $\obsope \etat ^t$ corresponds to the convolution over the core-mantle boundary 
of the true state with a Jacobi polynomial of degree $L$ \citep[paragraph 4.4.4]{backus1996}:
\begin{linenomath*}
\begin{eqnarray}
(\obsope \etat ^t)(\theta_o, \varphi_o) = 
\frac{L+1}{4 \pi} \int_{\theta=0}^{\pi} \int_{\varphi=0}^{2\pi} \etat ^t(\theta, \varphi) P_L^{(1,0)} (\cos \alpha) 
\sin \theta \mathrm{d}\theta\mathrm{d}\varphi, \label{jacobi}
\end{eqnarray}
\end{linenomath*}
in which $(\theta_o, \varphi_o)$ are the coordinates of the observation locations, 
$\alpha$ is the angular distance between the points $(\theta_o, \varphi_o)$ and $(\theta, \varphi)$, and 
$P_L^{(1,0)}$ is a Jacobi polynomial  \citep[p. 773]{abramowitz1965}. 
In the following experiments, we set $L=15$ and observations are made at a fixed temporal frequency.

Next, we start the assimilation with a different set of initial conditions, called initial guess, $\etat_0^g$. 
After a forward integration, the computed observable, $\obsope \dot{\magr}^f$ in section \ref{sectionsteady} and 
$\obsope \magr^f$ in section \ref{sectiontorsion}, 
is compared (over the entire time window) with the observations, 
and the discrepancy between the two gives the initial misfit (see equation~(\ref{obsmisfit})).
After assimilation, the decrease of the misfit and the relative difference between $\etat^t$ and $\etat^a$, in the $l^2$-sense, 
are used to assess the quality
of the recovered state.

\subsection{The kinematic core flow problem}
\label{sectionsteady}
In this section, a connection is made between core flow kinematic inversions and data assimilation. 
The steady flow hypothesis has been previously used by \cite{voorhies1985} and \cite{waddington1995} to break
the non-uniqueness of the kinematic inversion problem.
Here, we study the effect of a steady non-zonal and equatorially symmetric flow on the evolution of the radial 
magnetic field, and more particularly its secular variation $\dot{\magr}$, 
\begin{linenomath*}
\begin{eqnarray}
\dot{\magr} = - \divH  \left( \NZ{\velo} \magr \right). \label{eqsteady}
\end{eqnarray}
\end{linenomath*}
Symmetry with respect to the equator ensures uniqueness of the solution when $\magr $ and $\dot{\magr}$ are perfectly known.
The time scale characterizing this problem is the advection time, $t_{adv}$.
The non-zonal velocity effectively enters equation~(\ref{eqsteady}) via the non-zonal streamfunction 
$\Psi (s,\varphi)$ (see equation~(\ref{defpsi})), 
projected at the core-mantle boundary.
The state vector $\etat$ includes a parameter, the streamfunction $\Psi$, and a variable, 
the radial magnetic field $\magr$. Here, $\Psi$ is called a parameter
because it is steady in that experiment.
We seek the distribution of $\Psi (s, \varphi)$ which best explains the observed synthetic database of secular variation 
$\dot{\magrobs}$ at the top of the core.

The tangent linear form of (\ref{eqsteady}) is
\begin{linenomath*}
\begin{eqnarray}
\delta \dot{\magr} &=& {\sf P} \left(\delta \Psi, \magr \right) + {\sf Q} \left( \Psi, \delta \magr\right), \label{LINTANeqsteady} \\
\ddt \delta \magr &=& \delta \dot{\magr},  \label{LINTANeqsteady2}
\end{eqnarray}
\end{linenomath*}
where $\delta \Psi$, $\delta \magr$ and $\delta \dot{\magr}$ are the differentials of $\Psi$, $\magr$ and $\dot{\magr}$, respectively, 
and  $ {\sf P}$ and $ {\sf Q}$ 
the differentials of the right-hand side term of equation~(\ref{eqsteady}) with respect to $\Psi$ and $\magr$, respectively 
(they are developed in appendix~\ref{app_steady}).
Let us introduce  $\ADpsi, \ADBr$ and $ \dot{\magr}^T$  as the adjoint variables
of $\Psi$, $ \magr$ and  $\dot{\magr}$. The adjoint model of equations~(\ref{LINTANeqsteady}) and (\ref{LINTANeqsteady2}) is then 
\begin{linenomath*}
\begin{eqnarray}
\ADpsi  &=& \sum_{j=0}^T {\sf P}^T \left(\discrADBrdot, \discrmagr \right),\\
\ADBr &=& {\sf Q}^T \left(\dot{\magr}^T, \Psi \right),
\end{eqnarray}
\end{linenomath*}
in which  $ {\sf P}^T,  {\sf Q}^T$ are the adjoint functions 
of $ {\sf P}$ and $ {\sf Q}$  
(see detailed equations in appendix~\ref{app_steady}). 
The link to the observations has been obtained from equation~(\ref{obsmisfit}), 
and is computed at each time step as in equation~(\ref{eqadjoint}): 
\begin{linenomath*}
\begin{eqnarray}
\dot{\magr}^T &=& \alpha_H \left( \obsope ^T~ \obsope \dot{\magr} - \obsope ^T~ \dot{\magrobs} \right). \label{ADbr}
\end{eqnarray}
\end{linenomath*}

\begin{figure}
\centerline{\includegraphics[width=0.8\linewidth]{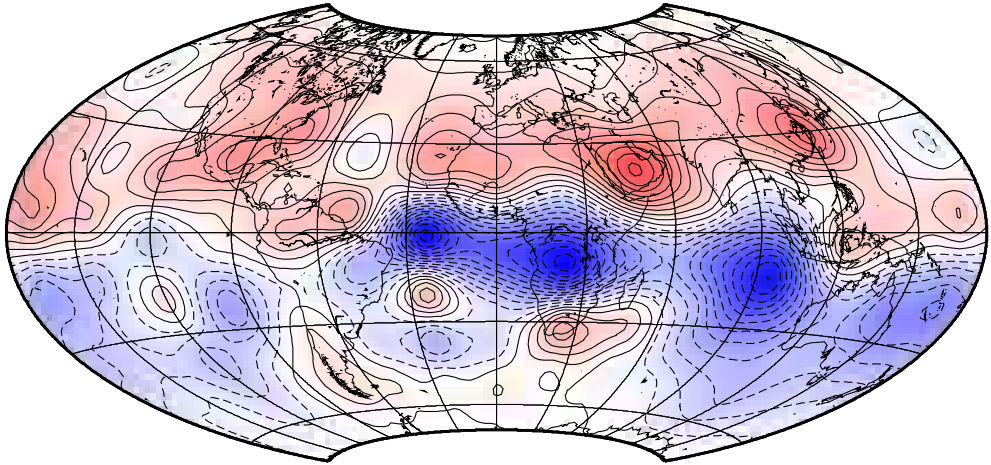}}
\caption{Map of the main field at epoch $2002$ truncated at spherical harmonic degree
and order $12$ modified from the CHAOS model  \citep{chaos06}.
True radial component of the magnetic field $\magr^t$ at initial time for the twin experiments. 
Contours are drawn each $0.5\magr^{rms}$ (solid (resp. dashed) for positive (resp. negative) values).
Note that the problem is linear in $\magr$. }
\label{br_initchaos}
\end{figure}

\begin{figure}
\centerline{\includegraphics[width=0.99\linewidth]{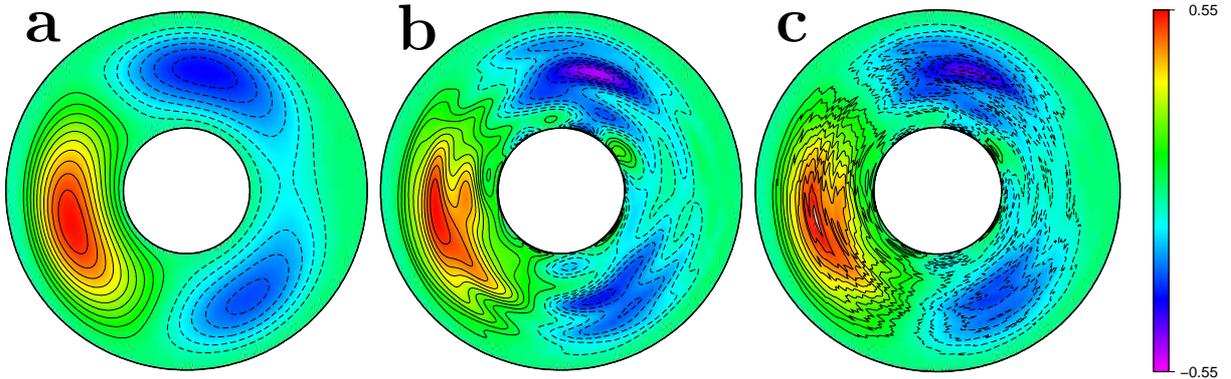}}
\caption{Maps in the equatorial plane $\Sigma$ of the steady streamfunction $\Psi$: a: true state, 
b: analyzed state with a single epoch inversion, c: analyzed state with $T=0.57~t_{adv}$.
Contours are drawn each $0.05$ (solid (resp. dashed) for positive (resp. negative) values). 
Extrema are $-0.33;0.54$ (a),  $-0.49;0.54$ (b) and  $-0.43;0.56$ (c).}
\label{resSteady}
\end{figure}

The trajectory of the true state is computed from the following set of initial conditions:
\begin{enumerate}
\item $\magr^t(\theta, \varphi,t\!=\!0)$ is obtained from the CHAOS main field model \citep{chaos06} for epoch $2002$, truncated at spherical harmonic degree 
and order $12$. It is taken outside the tangent cylinder and multiplied by a sine function of $\theta$ 
in order to have $\magr^t(\theta, \varphi,t\!=\!0)=0$ at the tangent cylinder (see Figure~\ref{br_initchaos}), 
\item $\Psi^t (s, \varphi)$ is shown in Figure~\ref{resSteady}a, it is the non-zonal part of an inverted flow from \cite{pais2004} truncated 
at degree and order $4$, and multiplied by $\cos ^2 \theta=(1-s^2)$ and a function of $s$, ($s-r_i$), 
in order to satisfy the flow boundary conditions at $s=r_i,r_o$. It is normalized in order to have a dimensionless rms velocity 
of order 1; the scaling $V_{adv}$ is such that 
$\int_0^{2\pi} \int_{\theta_c}^{\pi-\theta_c} (u_s^2+u_\varphi^2) \sin \theta\mathrm{d}\theta\mathrm{d}\varphi = 
V_{adv}^2 \int_0^{2\pi} \int_{\theta_c}^{\pi-\theta_c} \sin \theta \mathrm{d}\theta\mathrm{d}\varphi $, with $\theta_c=\mathrm{asin}(r_i/r_o)$.
\end{enumerate}

We consider perfect observations, setting $\obserr = {\bf 0}$ in equation~(\ref{phys2obs}). 
For the following simulations, the numerical time step is $6\times10^{-5}\ t_{adv}$ for 
integration times ranging from $0.03\ t_{adv}$ to $0.57 \ t_{adv}$.  
Other numerical parameters relative to the simulations are given in appendix~\ref{app_nummodel}.

In this problem of seeking a steady streamfunction that explains the observations, we want to show the 
benefit of including the temporal dimension (data assimilation) instead of relying on a single observation epoch, as it is the case 
for a standard kinematic inversion.

We first consider solutions obtained with only one observation epoch
and less observation locations ($N_\theta^O=50, N_\phi^O=11$) 
than grid points ($N_\theta=200, N_\phi=15$).
We start the assimilation with a first guess $\Psi^g$ corresponding to the minimal hypothesis: $\Psi^g=0$.
The solution we obtain gives us some information about what could be achieved within the  kinematic framework in that configuration.
Figure~\ref{resSteady}b shows in particular that the true state (Figure~\ref{resSteady}a) is not completely recovered, 
due to the truncation used in the construction of $\obsope$ and the limited number of observation locations, compared to the total number of grid points.

Using this solution obtained from a single epoch inversion as a reference solution, we can study the benefit we get resorting to a timeseries 
of observations, as opposed to a single snapshot. In other words, we investigate whether the issue of spatial subsampling can be partially 
fixed by considering 
the temporal dimension.
To that end, we do experiments with assimilation time windows ranging from $0.03\ t_{adv}$ to $0.57 \ t_{adv}$, at a given temporal frequency of 
observation  $f_{\obs}=100 \ t_{adv}^{-1}$, keeping the same number of virtual magnetometers.

\begin{figure}
\centerline{\includegraphics[clip=true,width=0.99\linewidth]{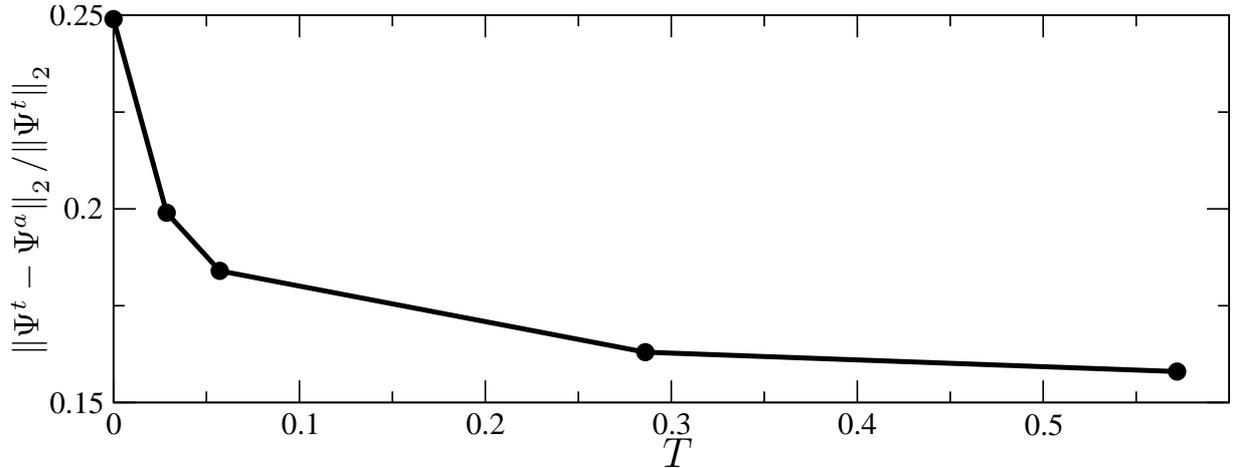}}
\caption{Effect of the assimilation time $T$ on the analyzed state at a fixed observation frequency $f_{\obs}=100\ t_{adv}^{-1}$, measured with 
the norm of the relative error 
$\lnorm{\Psi^t-\Psi^a} \left/ \lnorm{\Psi^t} \right.$. 
This norm is equal to one when the initial guess $\Psi^g=0$. 
The simulation at $T=0$ is referred in the text as the single epoch inversion.} 
\label{psiaVST}
\end{figure}

Results of a typical experiment are shown in Figure~\ref{resSteady}c, for which 
$T=0.57 \ t_{adv}$. 
The large scale pattern is retrieved, but the solution is polluted by small spatial scales 
(no extra smoothing term is added to the misfit function). 
We find, however, that the consideration of the temporal dimension improves the solution. 
Moreover, 
the distance between the true streamfunction and the retrieved streamfunction  becomes smaller when the 
assimilation time window is widened (see Figure \ref{psiaVST}).

As described by \citet[chapter 6]{evensen2007}, if one enlarges the width of the assimilation time window in a non-linear context, 
the misfit function presents 
more and more spikes and minima. A very good first guess is thus needed to converge to the global minimum.
To circumvent this issue, we decided to use the results obtained over short assimilation time windows as initial guesses for assimilation over 
longer time windows. That strategy is analogous to the approach used in the atmospheric variational assimilation community, which consists in 
solving a series of strong constraint inverse problems, defined for separate subintervals in time \citep[e.g.][]{evensen2007}.

\subsection{Forward and adjoint modeling of Alfv\'en torsional waves}
\label{sectiontorsion}

In non-rotating magnetized flows, classical Alfv\'en waves result from the balance between inertial and magnetic forces.

In the Earth's core, where the Coriolis force is large, \cite{braginsky1970} showed that a special class of Alfv\'en waves 
comes into play, in which only the component of the magnetic 
field normal to the axis of rotation, $\mags$, participates.
Associated motions are geostrophic; they are organized in axial cylinders about the axis of rotation, hence the name torsional oscillations. 
The period of torsional waves depends on the strength and distribution
of  $\mags$ inside the core.

In order to study these waves, one can consider a subset of the complete dynamical model. 
Since torsional waves are geostrophic and axisymmetric motions,
let us discard the non-zonal part of the flow and magnetic induction in equations~(\ref{eqvort}) to (\ref{eqbsbp}).

In addition to the vertical average, $\left\langle \cdot \right\rangle$, introduced in equation~(\ref{moyvert}), 
we now define the average  on a geostrophic cylinder, $\left\{ \cdot  \right\}$, by 
\begin{linenomath*}
\begin{eqnarray}
\left\{ X \right\}  \left( s\right) = \intp{\left\langle X \right\rangle}.
\end{eqnarray}
\end{linenomath*}
In the following, we will not indicate explicitly the dependence on $s$ of quantities 
averaged on a geostrophic cylinder.
Application of $\left\{ \cdot \right\}$ to equation~(\ref{eqzon}) yields 
\begin{linenomath*}
\begin{eqnarray}
\ddt \omeg =  \left(s^3h \right)^{-1} \dds \left( s^2h \axiBsBp  \right). \label{eqzon_torsion}
\end{eqnarray}
\end{linenomath*}
Similarly, the equations governing the evolution of magnetic quantities become
\begin{linenomath*}
\begin{eqnarray}
\ddt \axiBsBs &=& 0, \\
\ddt \axiBsBp &=& s \axiBsBs \dds \omeg.
\end{eqnarray}
\end{linenomath*}
Written in terms of the geostrophic angular velocity $\omeg$, the torsional wave equation is
\begin{linenomath*}
\begin{eqnarray}
  \ddt^2 \omeg  &=&  \left(s^3h \right)^{-1} \dds \left(s^3h \axiBsBs \dds \omeg \right).
\label{waveeq}
\end{eqnarray}
\end{linenomath*}
Equation~(\ref{waveeq}) can be transformed into a set of two first-order equations
\begin{linenomath*}
\begin{eqnarray}
\ddt \omeg  &=&  \left(s^3h \right)^{-1} \dds \tau, \\
\ddt \tau &=& s^3h  \axiBsBs \dds \omeg ,
\label{eqaxibs2}
\end{eqnarray}
\end{linenomath*}
in which $\tau=s^2h \axiBsBp$ is an auxiliary variable.

We have taken as boundary condition for the angular velocity:
\begin{linenomath*}
\begin{eqnarray}
\dds \omeg = 0, \mbox{\ \ at $s = r_o$.} 
\label{bcstorsion1}
\end{eqnarray}
\end{linenomath*}

Then, the boundary condition,
\begin{linenomath*}
\begin{eqnarray}
\dds \omeg = 0, \mbox{\ \ at $s = r_i$,} 
\label{bcstorsion2}
\end{eqnarray}
\end{linenomath*}
ensures the conservation of the total angular momentum carried by the fluid in
the computational domain. 

Projected at the CMB, those motions interact with $\magr$ through 
equation~(\ref{tracerbr}), which simplifies here into 
\begin{linenomath*}
\begin{eqnarray}
\ddt \magr &=& - \omeg \ddp \magr.
\end{eqnarray}
\end{linenomath*}
The system state 
gathers the geostrophic angular velocity $\omeg$, the variable $\tau$, 
the cylindrical average of the $s$-component of the magnetic field, $\axiBsBs$, 
and the radial component of the magnetic field $\magr$ at the CMB.

We define $\ADomeg,\ADtau, \ADBsBs, \ADBr$ as  the adjoint variables
of $\omeg,\tau, \axiBsBs, \magr$ respectively. The torsional oscillations adjoint model is
\begin{linenomath*}
\begin{eqnarray}
-\ddt \ADomeg  &=& \dds^T \left[ s^3h \axiBsBs \ADtau \right] - \ADBr \ddp \magr,  \label{torsionadX} \\
-\ddt \ADtau &=& \dds^T \left[ \left(s^3h \right)^{-1} \ADomeg \right], \\
\ADBsBs(s) &=& \sum_{j=0}^T s^3h \left( \dds  \omeg \right)_j  \ADtau_j  + \alpha_C \const \axiBsBs,  \label{torsionadK} \\
-\ddt \ADBr &=& - \ddp^T \left[\omeg \ADBr\right],
\end{eqnarray}
\end{linenomath*}
where $\dds^T$ and $\ddp^T$ are the adjoints of the differential operators 
$\dds$ and $\ddp$ (see appendix~\ref{app_torsion} for more details 
on the adjoint system).
The link to the constraint on $\axiBsBs$  has been obtained from equation~(\ref{obsmisfit}).
The model is completed by the information supplied by the observations, as in equation~(\ref{ADbr}).
The boundary conditions for the adjoint model are $\ADtau=0$, at both $s = r_i$ and $s = r_o$.
The temporal and spatial discretizations of this problem are described in appendix~\ref{app_nummodel}. 

\begin{figure}
\centerline{\includegraphics[width=0.99\linewidth]{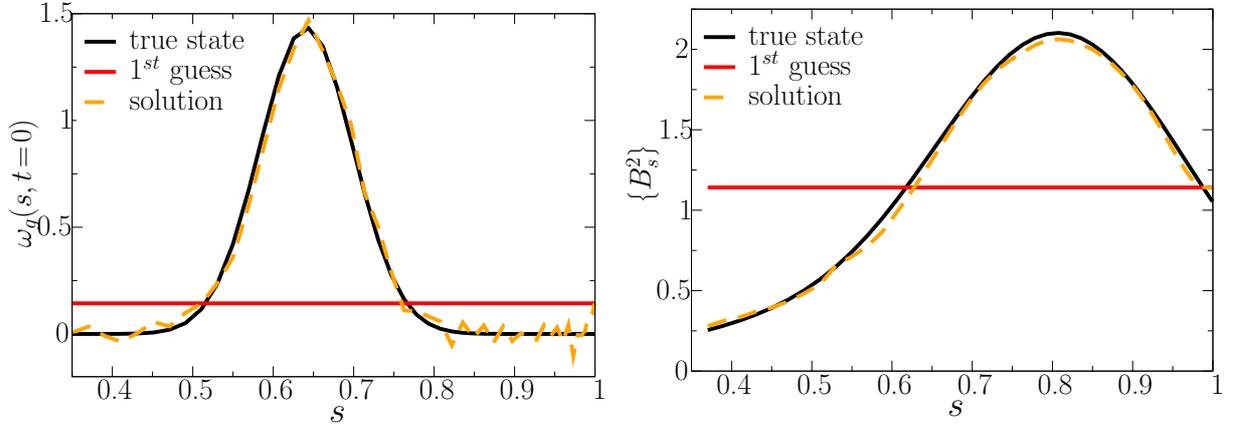}}
\caption{Torsional wave twin experiments results: true state (black), profile before assimilation (red) and solution after assimilation (dashed-orange) 
for $\omeg$ (left) and $\axiBsBs (s)$ (right). The parameters for that reference case are $\omegrms=0.34$ and $T=1.16\ T_A$.
Regularization has been added to the spatial derivative of $\axiBsBs$, of amplitude $\alpha_C=10^{-8}$ (See text for details).
}
\label{refcase1}
\end{figure}

For the experiments that follow, 
the set of initial profiles, which define the true state, is:
\begin{enumerate}
\item the same $\magr^t(\theta, \varphi,t=0)$ as in the kinematic core flow problem of section~\ref{sectionsteady},
\item a Gaussian function for the angular velocities: $\omeg^t(s,0)~=~\omegrms \exp \left[-\sigma_{\omega}^{-2}(s-s_{\omega})^2 \right] $, 
with $\sigma_{\omega}^{-2}=150$ and $s_{\omega}=0.65$; it satisfies 
the boundary conditions (\ref{bcstorsion1}) and (\ref{bcstorsion2}), its amplitude being scaled by $\omegrms$ (discussed hereafter),
\item $\tau^t(s,0) = 0$,
\item  an arbitrary function $\axiBsBs^t$ (see the black curve in Figure~\ref{refcase1}, right) given by:

$\axiBsBs^t(s)=c_1+c_2\sin\left(\pi/2-L\right)+c_3\exp\left[-\sigma_{B}^{-2}\left(s-s_{B}\right)^2\right]$, with 
$c_1=0.1$, $c_2=0.02$, $c_3=1$, $\sigma_{B}^{-2}=20$, $s_{B}=0.8$ and $L=14s$.  
It is normalized in order to have a dimensionless rms magnetic field of unity inside the core; the scaling $B_0$ is such that
$\int_0^{2\pi} \int_{r_i}^{r_o} B_s^2 \mathrm{d}s\mathrm{d}\varphi = B_0^2 \int_0^{2\pi} \int_{r_i}^{r_o} \mathrm{d}s\mathrm{d}\varphi $.
\end{enumerate}

As the velocity has been scaled by the Alfv\'en velocity,  $\omegrms$ is the ratio between the Alfv\'en time and the advection time, 
$\omegrms=T_A/t_{adv}$. 

Assimilation is performed on both $\omeg$ and $\axiBsBs$. 
We seek the steady profile of $\axiBsBs$ and the initial profile of $\omeg$ which best explain the synthetic database of $\magrobs$.
Our first guess consists of flat profiles: $\omeg^g(s,0) = 0.1\omegrms$ and $\axiBsBs^g = 0.6$  
(see the red curves in Figure~\ref{refcase1}). 

\begin{figure}
\centerline{\includegraphics[clip=true,width=0.55\linewidth]{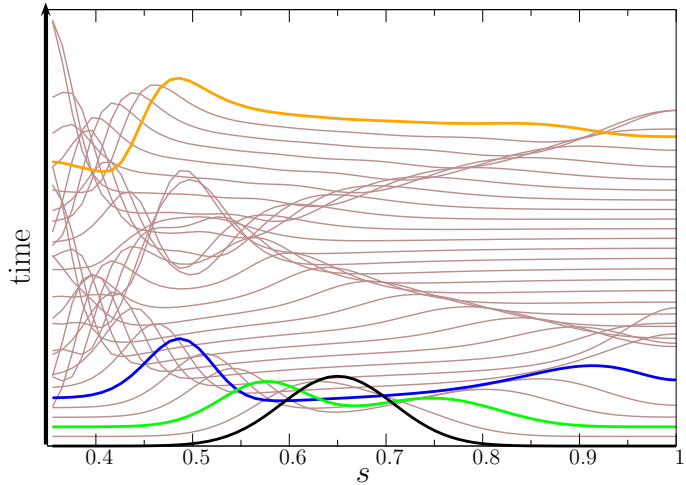}}
\caption{Torsional wave twin experiments results: 
successive profiles of the angular velocity $\omeg$, showing the 
propagation of a torsional wave in the computational domain $s \in \left[0.35,1 \right]$ during $1.16$ Alfv\'en time $T_A$. 
Initial condition (black curve) for $\omeg$, 
and snapshots, at $0.12$ (green), $0.18$ (blue) and $1.16\ T_A$ (orange).}
\label{torsionforward}
\end{figure}

\begin{figure}
\centerline{\includegraphics[width=0.99\linewidth]{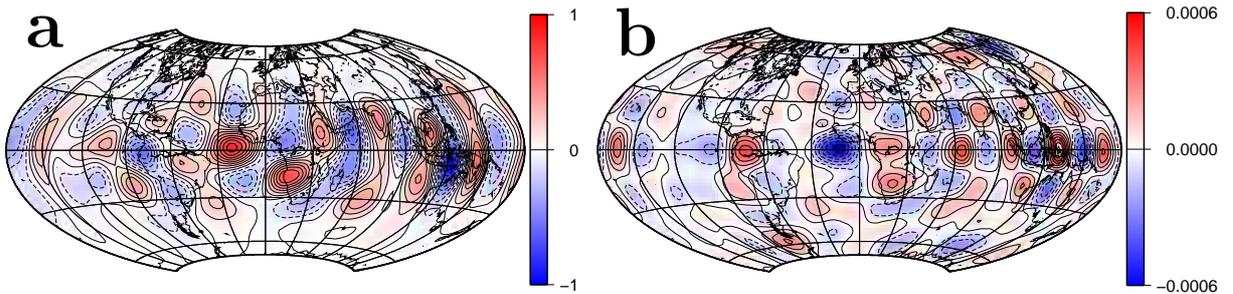}}
\caption{Relative difference between observed and computed $\obsope \magr$  at final time,  
$\left[\magrobs(T) - \obsope \magr^f(T)\right]/\lnorm{\magrobs(T)}$ before 
assimilation ({\bf a}) and after assimilation ({\bf b}) for the reference case (same parameters as in Figure \ref{refcase1}). 
Contours are drawn each $0.1$ ({\bf a}) and $10^{-1}$ ({\bf b}) (solid (resp. dashed) for positive (resp. negative) values).
Extrema are $-0.94;0.86$ ({\bf a}) and $[-6.4;7.2]\!\times\!10^{-4}$ ({\bf b}).}
\label{refcase2}
\end{figure}

\begin{figure}
\centerline{\includegraphics[width=0.99\linewidth]{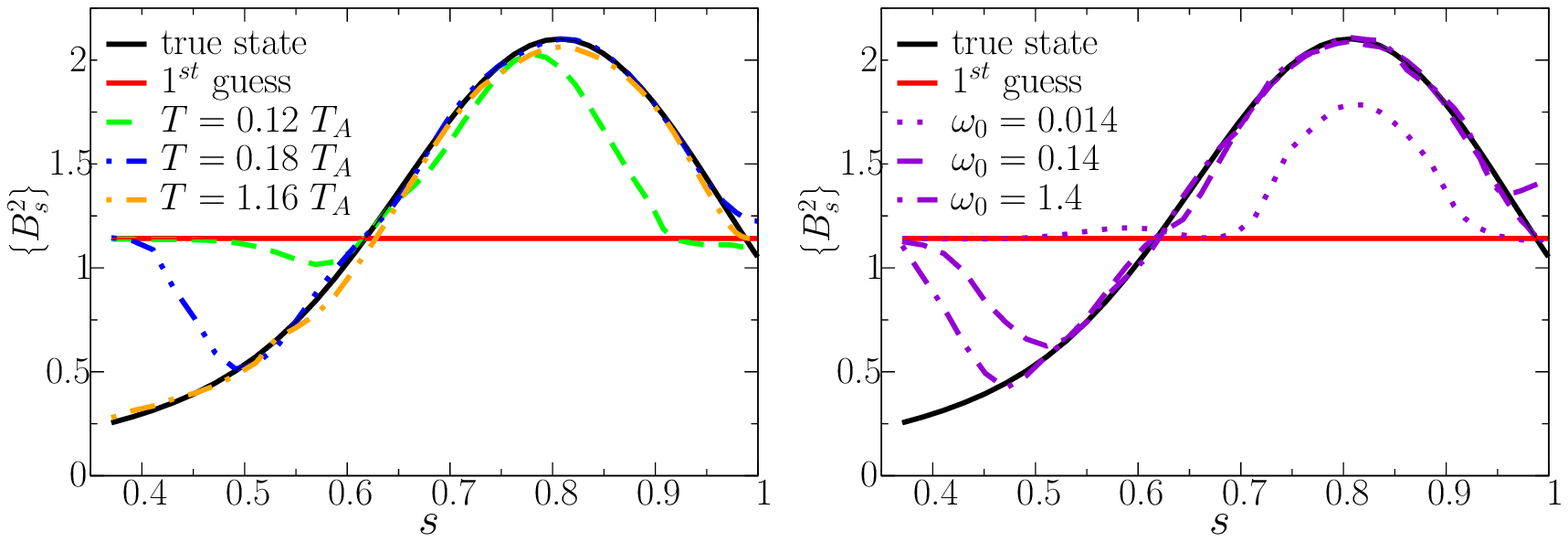}}
\caption{Effect on the analysis of the assimilation time window of width $T$ (left) and of the amplitude of the initial pulse $\omegrms$ (right).
Left: true state (black), profile before assimilation (red) 
and solution after assimilation (dashed-green, dashed-double-dotted-blue, dashed-dotted-orange)
for $\axiBsBs (s)$, for different values of $T$ ($0.12, 0.18$ and $1.16\ T_A$ respectively) for a fixed $\omegrms=0.34$.
Right: true state (black), profile before assimilation (red) and solution after assimilation (violet) for $\axiBsBs (s)$. 
Dotted, dashed and 
dashed-dotted violet curves are obtained for different values of the amplitude of the initial pulse $\omegrms$ ($0.014, 0.14$ and $1.4$ respectively), 
keeping $T$ fixed, equal to $0.18\ T_A$.  }
\label{effetTurms}
\end{figure}

We show here experiments with a fixed frequency of observations $f_{\obs}~=~20\ T_A^{-1}$ and as many observations locations as grid points.
The observations are blurred by the averaging kernel (equation~(\ref{jacobi})), which causes errors. 
Consequently, the analysis can develop small scales, which are not very well constrained by the observations. 
We choose to reduce the complexity of the solution by adding a smoothing term to the cost function, taking 
$\alpha_C=10^{-8}$ 
in equation~(\ref{obsmisfit}). Here we penalize only the strong spatial gradients of $\axiBsBs$.
The reference case (with $\omegrms=0.34$ and $T=1.16\ T_A$, see Figures~\ref{refcase1} and \ref{refcase2}) shows 
that both the angular velocity and interior magnetic field are well 
recovered. 
As shown in Figure~\ref{refcase2}, the error field is substantially weaker after assimilation.

On a technical note, 
the M1QN3 algorithm \citep{gilbert89}, used in the optimization loop, stops in that case when the 
initial misfit is divided by a factor of 
$4 \times 10^5$, which is reached in $214$ iterations.

In order to assess the effect of the width of the assimilation window on the retrieved state variables, we vary 
the assimilation time $T$ between $0.12$ and $1.16\ T_A$, keeping $\omegrms$ constant, equal to $0.34$ as above. 
The geostrophic angular velocity is in all cases completely recovered (not shown) 
with similar spurious oscillations as in Figure~\ref{refcase1} (left)
near the outer boundary.
On the other hand, the area over which $\axiBsBs$ is correctly retrieved increases with  $T$,
indicating that the assimilated area is controlled by the distance over which the initial pulse has propagated (Figure~\ref{effetTurms}, left). 
Figure~\ref{torsionforward} shows that, if $T=1.16\ T_A$, the wave has enough time to explore 
the whole domain. On the contrary, if $T=0.12$ or $0.18 \ T_A$, a lesser portion of the 
domain is sampled by the wave, over which $\axiBsBs$ has been effectively 
retrieved. 
The angular velocity is better recovered than $\axiBsBs$ because it is directly connected 
to the observations, as opposed to $\axiBsBs$. 
That can be seen in the adjoint equations: $\ADomeg$ (equation~(\ref{torsionadX})) depends on $\ADBr$ that contains
$\magrobs$, the observed quantity, whereas $\ADBsBs$ (equation~(\ref{torsionadK})) is only directly connected to $\ADtau$. 
In turn, $\ADtau$ sees $\ADomeg$,
which is ultimately linked to the observations.

For a fixed $T$, the dependence on the amplitude $\omegrms$ has been studied (Figure~\ref{effetTurms}, right). 
For $\omegrms=0.014$, $\omeg$ and $\axiBsBs$ are not well recovered.
Starting from $\omegrms=0.14$, however, increasing  $\omegrms$ by more than one order of magnitude 
has little effect on the retrieval of $\axiBsBs$ and no effect at all on $\omeg^a$.

Let us stress  that the convergence of the calculations presented here is also sensitive 
to the profile of the initial condition (for both true state and first guess),  
and to the amount of measurements, as in the steady case of section~\ref{sectionsteady}.
Moreover, we have observed in other instances (not shown) that convergence is sped up 
if the acceleration $\tau$ is not zero at initial time.
More generally, that particular example shows that the success of assimilation is controlled 
by the intrinsic dynamics of the system under study, 
as well as a good guess of its state.

\begin{figure}
\centerline{\includegraphics[width=0.99\linewidth]{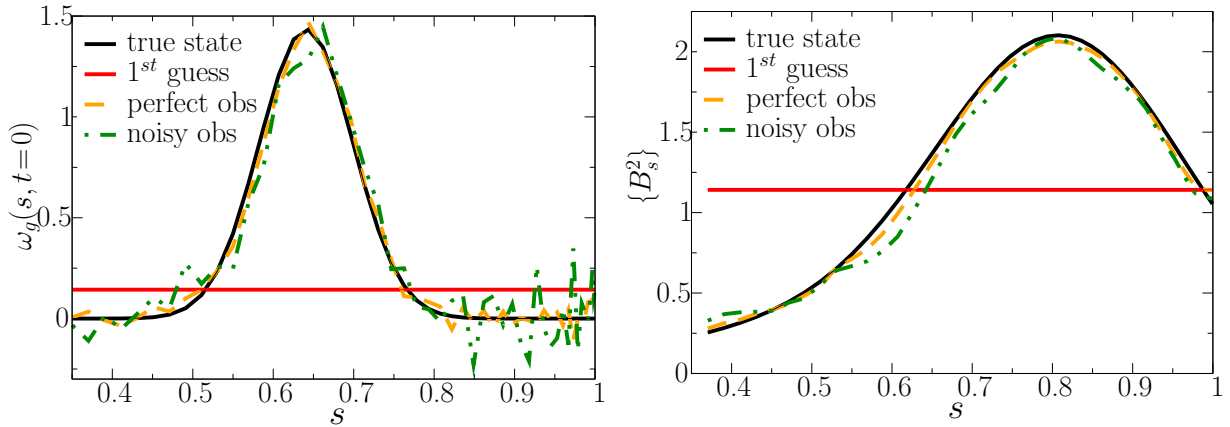}}\caption{Torsional wave twin experiments results obtained with noisy data: 
True profiles (black), profile before assimilation (red) 
and solution after assimilation (orange, green) for $\omeg$ (left) and $\axiBsBs$ (right). 
Dashed-orange curves have been computed with perfect observations, dashed-dotted-green curves considering centered, 
normally distributed Gaussian observation errors  of standard deviation $10^{-1}\magr^{rms}$ added to the database.
Regularization has been added on the spatial derivative of $\axiBsBs$, of amplitude $\alpha_C=10^{-8}$.
The assimilation window width is $T=1.16\ T_A$, and $\omegrms=0.34$.}
\label{result_noise}
\end{figure}

Until now, we have assumed perfect observations: $\obserr = {\bf 0}$ in equation~(\ref{phys2obs}). 
In the prospect of future applications, observation error should be considered. 
In the next experiments, observations contaminated by errors are assimilated.
Centered, normally distributed Gaussian  observation errors of standard deviation $10^{-1} \magr^{rms}$ are added to the previous database.
That experiment is carried out with an assimilation window width of $1.16\ T_A$.
Even with a database contaminated by observation errors, it is still possible to recover the shape and strength of the true state 
(see Figure~\ref{result_noise}).
$\axiBsBs$ seems less sensitive to observation errors than $\omeg$: we still have an extra penalty term in the misfit function as 
in the perfect observations case, but the overall effect of small scales is actually to degrade the solution for both fields.

For completeness, we have also decreased the temporal frequency of observation and observed that the recovering of the true state was possible 
provided that the frequency of observation, $f_{\obs}$, was greater than $3\ T_A^{-1}$. 

\section{Discussion}

We have derived a quasi-geostrophic model of core dynamics, which aims at describing core processes
on geomagnetic secular variation timescales.
Under the quasi-geostrophic assumption, the magnetohydrodynamics takes place in the equatorial plane 
 and is written outside the tangent cylinder.
The flow is defined by its zonal velocity and non-zonal axial vorticity.
The magnetic induction appears through $z$-averaged quadratic magnetic quantities, 
while we assume that the magnitude of the magnetic field
at the core surface is smaller than in the core interior.
In addition, the equatorial flow is projected on the core-mantle boundary. It interacts with the
magnetic field at the core surface, through the radial component of the magnetic induction equation, in the frozen-flux
approximation. That part of the model connects the dynamics and the observed secular variation, with
the radial component of the magnetic field acting as a passive tracer.
We have resorted to variational data assimilation to construct formally the relationship between model predictions and observations.
The use of an imperfect observation operator mimics our truncated vision of the reality.
We have extensively tested the variational data assimilation algorithm with twin experiments owing to the 
non-prohibitive numerical costs of the computations.
Assimilation was controlled by the initial state and possibly some static model parameters.

Let us stress some important results from our numerical simulations.
In our time-dependent framework, we have found that increasing the
time window width $T$ always improves the solution. In the steady core
flow
experiment case, that property has proven useful in constraining
intermediate flow length scales otherwise unconstrained by a sparse
distribution of observations. Here, the benefit originates from the
dynamical relationship which exists between successive observations through the radial component of the magnetic induction equation.
In the torsional oscillation case (an illustration of the dynamical
model presented in this study), the same property allowed us to
retrieve the $z$-averaged quadratic product of $B_s$ (which is only
remotely linked to the observed quantity) over
the entire domain provided that $T$ was large enough for the wave to propagate
over (and effectively sample) the whole domain.
Interestingly, it has not been necessary to include a dissipation term in the forward model. 
We have investigated the sensitivity of the solution to the frequency of observation in the presence of observation errors.
Adding an extra smoothing term to the cost function proved an efficient way to produce analyses  
with a moderate level of complexity.

From the geophysical point of view, with a typical estimate 
of the magnetic field strength in the core interior, $2$~mT, one Alfv\'en time, $T_A$, amounts to $6$ years.
We have varied the frequency of observation from $2.5$ to $20\ T_A^{-1}$, $3\ T_A^{-1}$ appearing as a minimum to recover the fields, which  
represents a two-year interval between observations.  
Therefore, we expect that we may be able to resolve properly torsional waves using the last $10$ years of satellite measurements, since 
the most recent magnetic field models have a resolution of a fraction of year \citep{olsen2008}.

\cite{pais2008} and \cite{gillet2009} have recently used magnetic field models obtained from satellite data in kinematic 
inversions of quasi-geostrophic core flows. Their calculated core flows are dominated by a giant retrograde gyre.
 \cite{gillet2009} suspect that the weaker momentum of the gyre for the period $1960$-$1980$, compared to the period $1990$-$2008$, 
is an artifact produced by the lesser data quality before $1980$.
Resorting to a toy model, \cite{fournier07} have demonstrated that the benefit due to the existence of a denser network 
at the end of an assimilation window, is manifest over a substantial part of the window, 
thanks to the variational data assimilation approach. 
In other words, the recent high quality of observatory and satellite measurements can be in principle backward propagated 
in order to reassimilate historical data series.
It is then possible to imagine that 
the refined series could contribute to a more precise description of both small and large scales of the fluid circulation in the core.

The interplay between magnetic and rotation forces in a two-dimensional model has been investigated in other contexts. 
\cite{tobias2007} have recently studied a local two-dimensional $\beta$-plane numerical model to show the impact of a weak 
large-scale magnetic field
on the dynamics of the solar tachocline.
Instead of using quadratic products of the magnetic field as variables, 
they have written the magnetic field as a function of a unique scalar potential  $A$: 
\begin{linenomath*}
\begin{eqnarray}
\magn (s, \varphi) =  \rot \left[A (s, \varphi)\vectz \right].
\label{scalarpot}
\end{eqnarray}
\end{linenomath*}
Then, the magnetic term in the vorticity equation becomes $\left[\rot \left(A\vectz\right) \right]~\cdot~\boldnabla \left[ \laplE A \right] $
(compare with the right-hand side term of equation~(\ref{eqvort})), and the induction equation is
\begin{linenomath*}
\begin{eqnarray}
\ddt A = - \velo \cdot \boldnabla A + S^{-1} \laplE A.
\label{scalarind}
\end{eqnarray}
\end{linenomath*}
The ansatz (\ref{scalarpot}) is restrictive, 
as axial invariance of the magnetic field is assumed.
It enables the inclusion of magnetic diffusion, the effect of which cannot 
be rigorously introduced in the set of equations~(\ref{eqbs2}) to (\ref{eqbsbp}).
The model given by equations~(\ref{scalarpot}) and (\ref{scalarind}) is also attractive, since it is still 
able to describe a variety of physical situations.
As an example, \cite{diamond2005} mention the transition from two-dimensional magnetohydrodynamic turbulence at small 
length scales, to turbulence controlled by Rossby wave interactions at larger length scales.
Solutions of (\ref{scalarpot}) and (\ref{scalarind}) (without the diffusion term) are also solutions permitted by our equations~ 
(\ref{eqvort}) to (\ref{eqbsbp}).
Investigating that simplified set of equations thus appears as an appealing intermediate step 
before the actual implementation of the less restrictive equations based upon quadratic magnetic quantities.

\myspace

{\bf Acknowledgments }

We thank the reviewers, A. Tangborn and A. De Santis, for their constructive comments on the manuscript.
We thank E. Cosme, C. Finlay, M. Nodet and O. Talagrand for fruitful discussions. 
We also thank N. Gillet, N. Schaeffer and A. Pais for many exchanges at different stages of this work. 
The manuscript has benefited  from the suggestions received from C. Finlay and N. Gillet who read an earlier draft of the manuscript. 
A. Pais has provided a core flow model \citep{pais2004}. 
The M1QN3 routine has been provided by J. Gilbert and C. Lemar\'echal \citep{gilbert89}.
Maps have been produced with GMT free software \citep{wesselGMT1991}.

This work has been supported by a grant from the Agence
Nationale de la Recherche (“white” research program VS-QG, grant
reference BLAN06-2 155316) and by INSU, under the LEFE\_ASSIM program (Les Enveloppes Fluides et l'Environnement volet Assimilation).  


\myspace

\begin{appendix}
\section{A first-order variant of the induction equation}
\label{variante}
To write equations~(\ref{eqbs2}) to (\ref{eqbsbp}), we have retained only the zeroth-order part of the flow. 
We may wish to take into account, in these equations, the $z$-component of the flow that enters in the Coriolis term in 
equation~(\ref{eqvort}) and in the induction equation at the core-mantle boundary (equation~(\ref{tracerbr})).

Thus, in order to ensure incompressibility, we define

\beg
{\NZE{\velo}} (s, \varphi) = \gamma(s) \left[ \rot \Psi (s, \varphi)\vectz \right];
\fin
the continuity equation for ${\NZE{\velo}} + \velz^1 \vectz$ yields $\gamma(s)=h^{-1}(s)$.

The non-zonal vorticity field $\vort$ is then defined by $\mathbf \vort = \rot {\NZE{\velo}}$, and its vertical component is
\beg
\vortz (s, \varphi)= - \laplE \left[h^{-1}(s) \Psi(s, \varphi) \right].
\fin

Finally, the set of equations~(\ref{eqbs2}) to (\ref{eqbsbp}) becomes 
\beg
\ddt \BsBs &=& -\left[ \velo \cdot \nabla \right]\BsBs - 2 s^{-1}  \BsBs \vels - 2 s^{-1} \BsBs \ddp \velp + 2s^{-1}  \BsBp \ddp \vels, \\
\ddt \BpBp &=& -\left[ \velo \cdot \nabla \right]\BpBp - 2 \BpBp \dds \vels + 2 s \BsBp \dds \left(s^{-1} \velp \right), \\
\ddt \BsBp &=& -\left[ \velo \cdot \nabla \right] \BsBp + s \BsBs \dds \left(s^{-1} \velp \right)  + s^{-1} \BpBp \ddp \vels \nonumber \\
&& - \left[ \nabla _E \cdot \velo \right] \BsBp.
\fin

\section{Numerical model}
\label{app_nummodel}

\begin{figure}
\centerline{\includegraphics[clip=true,width=0.3\linewidth]{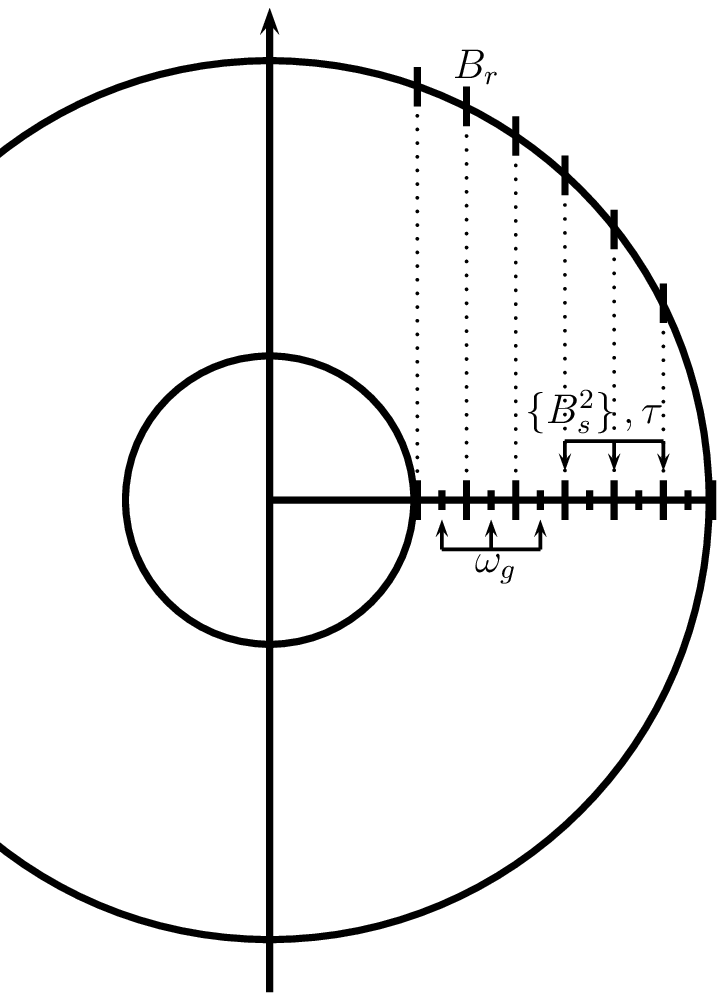}}
\caption{Sketch of a regular staggered radial grid in the equatorial plane, and its projection on the CMB.}
\label{discret}
\end{figure}

Fields are discretized in radius on a (possibly irregular) staggered grid (see Figure~\ref{discret}), 
$s = i_s \Delta_s(s); i_s \in \left[0,N_s\right]$; 
$\omeg(i_s + 1/2, j)$ and $\tau (i_s, j)$. $\Psi$, $ \axiBsBs$ and $\tau$ are calculated 
on the same spatial grid  (note that $ \axiBsBs$ is not defined on the endpoints). 
The latitudinal part of $\magr$ is discretized on a meridian and every grid point is mapped on the CMB from the grid point on $\Sigma$, 
except at the equator (see Figure~\ref{discret}); thus
$\theta = i_\theta \Delta_\theta(\theta) ; i_\theta  \in \left[0, 2N_s-1\right]$.

The azimuthal part of non-zonal fields, $\Psi$ and $\magr$ is  expanded in Fourier series:
\beg
\Psi(s, \varphi, t)&=& \sum_{m=0}^{m_{max}} \left[a_m(s, t) \cos \left(m \varphi \right) + b_m(s, t) \sin \left(m \varphi \right) \right], \\
\magr(\theta, \varphi, t)&=& \sum_{m=0}^{m_{max}} \left[c_m(\theta, t) \cos \left(m \varphi \right) + d_m(\theta, t) \sin \left(m \varphi \right) \right],
\fin
in which the number of azimuthal Fourier mode, $m_{max}$, is related to the  number of equidistant grid points in longitude, 
$N_\varphi$: $m_{max}=\left(N_\varphi-1\right)/2$.

The time step being $\Delta_t$, time is discretized using finite differences $t = j \Delta_t; j \in \left[0,N_t\right]$.

Spatial derivatives are computed with a finite difference scheme, 
except for the longitudinal derivatives for which we use the Fast Fourier Transform 
in order to compute them in spectral space.

For the simulations, the cylindrical radius is 
discretized in $N_s=100$ grid points (including the boundaries), the CMB in $N_\theta=200$ grid points 
in latitude and $N_\varphi = 33$ (unless otherwise specified) grid points in longitude.

\section{Steady non-zonal flow model}
\label{app_steady}
We define the streamfunction at the top of the core, $\Psi_o$, as $\Psi_o = \stotheta \Psi$, 
where subscript $o$ refers to the outer boundary.  
Let $ \stotheta$ be the operator which projects a vector from the equatorial plane to the top of the core, and let $\thetatos$ be its  transpose.
The forward model is the radial component of the induction equation at the top of the core,
\beg 
\dot{ \magr} &=& - \divH  \left( \NZ{\velo} \magr \right), \\
&=& \left(\sin\theta \cos\theta \right)^{-1} \left[ \ddtheta \Psi_o \ddp \magr 
-  \ddp  \Psi_o \ddtheta \magr \right]
+  \left(\cos \theta \right)^{-2} \Psi_o \ddp \magr.
\fin
The tangent linear equation is
\beg
\delta \dot{\magr} &=&
 \left(\sin\theta \cos\theta \right)^{-1} \left[ \ddtheta \Psi_o \ddp \delta \magr 
-  \ddp  \Psi_o \ddtheta \delta \magr \right]
+  \left(\cos \theta \right)^{-2} \Psi_o \ddp \delta \magr \\ \nonumber
& +& \left(\sin\theta \cos\theta \right)^{-1} \left[ \ddtheta \delta \Psi_o \ddp \magr 
-  \ddp  \delta \Psi_o \ddtheta \magr \right]
+  \left(\cos \theta \right)^{-2} \delta \Psi_o \ddp \magr, \\
\ddt \delta \magr &=& \delta \dot{\magr}. 
\fin
We introduce the adjoint variables $\ADpsi$, $\ADBr$  and $ \dot{\magr}^T$ for $\Psi$, $\magr$  and  $\dot{\magr}$ respectively.
The adjoint model is
\beg
\ADpsi&=& \thetatos \sum_{j=0}^T \left\{ 
\ddtheta^T \left[ \left(\sin\theta \cos\theta \right)^{-1}  \ddp \discrmagr \discrADBrdot  \right]
- \ddp^T \left[ \left(\sin\theta \cos\theta \right)^{-1}  \ddtheta \discrmagr \discrADBrdot  \right] \right. \\ \nonumber
&& \left. + \left(\cos \theta \right)^{-2} \ddp \discrmagr \discrADBrdot
\right\}\cos^2 \theta, \\
\ADBr &=& 
\ddp^T  \left[ \left(\sin\theta \cos\theta \right)^{-1}   \ddtheta \Psi_o  \dot{\magr}^T  \right]
- \ddtheta^T \left[ \left(\sin\theta \cos\theta \right)^{-1}   \ddp \Psi_o  \dot{\magr}^T  \right] \\ \nonumber
&&
+ \ddp^T \left[ \left(\cos \theta \right)^{-2} \Psi_o   \dot{\magr}^T \right], \\ \nonumber
 \dot{\magr}^T&=&-\ddt \ADBr, 
\fin
where $\ddp^T$ and $\ddtheta^T$ are the adjoints of $\ddp$ and $\ddtheta^T$,  and $-\ddt$ indicates that the integration 
is performed backward in time.

\section{Alfv\'en waves model}
\label{app_torsion}
The forward equations are:
\beg
\ddt \omeg  &=&  \left[s^3h \right]^{-1} \dds \tau, \\
\ddt \tau &=& s^3h  \axiBsBs \dds \omeg, \\
\ddt \magr &=& - \stotheta \Ztotot \omeg \ddp \magr,
\fin
where $\Ztotot$ transforms a one-dimensional zonal vector
into a two-dimensional one, by duplicating it in each meridional plane. Let $\tottoZ$ be its transpose.
The forward model is completed by the boundary conditions  (\ref{bcstorsion1}) and (\ref{bcstorsion2}).

We define the adjoint variables $\ADomeg,\ADtau, \ADBsBs, \ADBr$ 
for $\omeg,\tau, \axiBsBs, \magr$ respectively. The adjoint model is
\beg
\ddt \ADomeg  &=& \dds^T \left[ s^3h \axiBsBs \ADtau \right] - \tottoZ \thetatos \left[ \ddp \magr \ADBr \right] , \\ 
\ddt \ADtau &=& \dds^T \left[ \left(s^3h \right)^{-1} \ADomeg \right], \\
F^T(s) &=& \sum_j s^3h \ADtau_j \axiBsBs \left( \dds \omeg\right)_j + \alpha_C \const \axiBsBs, \label{torsionadF} \\
\ddt \ADBr &=& - \ddp^T \left[\stotheta \Ztotot \omeg \ADBr\right],
\fin
where $\dds^T$ is the adjoint of the operator $\dds$ and the term in $\alpha_C$ corresponds to the extra penalty term in the misfit function 
(see also equation \ref{regularizationterm}); $\const=\dds^T\dds$ in the experiments.
In order to enforce its positivity during the optimization phase, $\axiBsBs$ is rather written
$\axiBsBs = \exp \left[ F(s)\right]$,  
with $F \in \mathbb{R} $  and $F^T$ computed as indicated in equation~(\ref{torsionadF}).

\end{appendix}

\clearpage 


\end{document}